\documentclass[12pt]{iopart}
\usepackage{amssymb,bm}
\usepackage{graphicx}
\usepackage{amsfonts}
\usepackage{amssymb}
\usepackage{bbm}
\usepackage{enumitem}

\usepackage{graphicx}
\usepackage{subfigure}
\usepackage{color}

\newcommand{\bra}[1]{\langle #1 |}
\newcommand{\ket}[1]{| #1 \rangle}

\newcommand{\eqref}[1]{(\ref{#1})} 

\newcommand{\binom}[2]{{{#1}\choose{#2}}}

\usepackage{subfigure}
\usepackage{color}

\usepackage{graphicx}
\usepackage{bbm}
\usepackage{enumitem}

\usepackage[usenames,svgnames]{xcolor}
\usepackage{url}
\usepackage{umoline}
\usepackage[hyperindex,breaklinks]{hyperref}
\hypersetup{
     colorlinks=true,       		
     linkcolor=Salmon,          	
     citecolor=blue,            
     filecolor=blue,      		
     urlcolor=cyan,           	
 }

\newtheorem{theorem}{Theorem}
\newtheorem{lemma}{Lemma}

\newtheorem{definition}[theorem]{Definition}

\newtheorem{corol}[theorem]{Corollary}  
   
\newcommand{\ignore}[1]{{}}

\newcommand{\Oorderof}{\mathcal{O}}
\newcommand{\orderof}[1]{\Oorderof(#1)}

\newcommand{\Prod}{{\rm Prod}}
\newcommand{\co}{{\rm c}}

\newcommand{\ad}{{\rm ad}}
\newcommand{\dist}{{\rm dist}}
\newcommand{\diam}{{\rm diam}}
\newcommand{\for}{ \quad {\rm for} \quad}

\newcommand{\loc}{{\rm loc}}

\begin{document}

\title[]{Connecting probability distributions of different operators and generalization of the Chernoff-Hoeffding inequality}

\author{Tomotaka Kuwahara}

\address{WPI, Advanced Institute for Materials Research, Tohoku University, Sendai 980-8577, Japan}

\ead{tomotaka.kuwahara.e8@tohoku.ac.jp}


\begin{abstract}
This work is devoted to explore fundamental aspects of the spectral properties of few-body general operators. 
We first consider the following question: when we know the probability distributions of  a set of observables, what can we way on the probability distribution of the summation of them?
In considering arbitrary operators, we could not obtain a useful information over third order moment, while under the assumption of 
\textit{the few-body operators}, we can rigorously prove a much stronger bound on the moment generating function for arbitrary quantum states.
Second, by the use of this bound, we generalize the Chernoff inequality (or the Hoeffding inequality), which characterizes the asymptotic decay of the probability distribution for the product states by the Gaussian decay.  
In the present form, the Chernoff inequality can be applied to a summation of independent local observables (e.g., single-site operators). 
We extend the range of application of the Chernoff inequality to the generic few-body observables. 

\end{abstract}

\maketitle

\section{Introduction}
\label{sec:introduction}


In quantum mechanics, one of the most fundamental problems is to obtain probability distributions of observables, 
or equivalently to determine the parameters to describe a quantum state. 
Usually, in condensed matter systems, each of the spins (or particles) is not independent of each other, and hence 
such an analysis is generally quite a challenging problem~\cite{ref:KempeKitaevRegev} even for very simple quantum states~\cite{0034-4885-75-2-022001}.
The question is deeply related to understand the entanglement structure of states~\cite{ref:AL-rev,ref:Hastings2007-1D}.
Here, we mean the entanglement by the quantum correlation which comes from the superposition principle; for example, 
the product state like $\ket{0}^{\otimes N}$ has no entanglement. 
Usually, we can treat only slightly-entangled-states class~\cite{perez2006matrix,verstraete2008matrix,eisert2013entanglement}, because the number of parameters to describe generic quantum states increases exponentially with the system size. 
This in turn casts the problem of how to characterize slightly entangled states. 
In general, the complete analysis of the entanglement is extremely difficult for over four qubits states~\cite{RevModPhys.81.865},  and hence we need to characterize it in more coarse-grained manners.

\begin{figure}
\centering
\includegraphics[clip, scale=0.6]{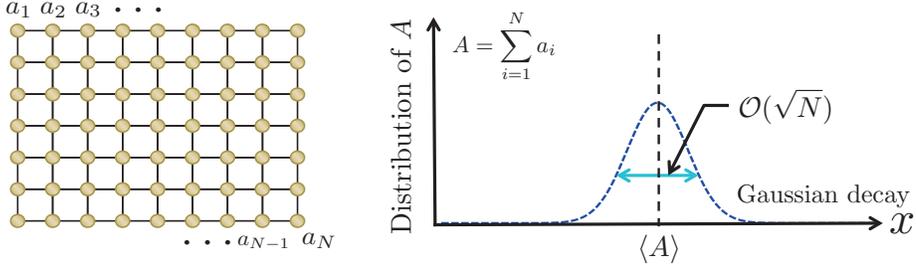}
\caption{Schematic picture of the Chernoff inequality. Let us consider a product state $\rho_{\Prod}$ and an operator $A$ which is given by the summation of single-site operators $\{a_i\}_{i=1}^N$ with $\|a_i\|=1$ for $i=1,2,\ldots,N$. The Chernoff inequality ensures that the probability distribution of $A$ for $\rho_{\Prod}$ decays faster than the Gaussian decay with the variance $\orderof{N}$. 
}
\label{fig:Chernoff_ineq}
\end{figure}

For the purpose, we here focus on the fact that in slightly entangled states the spins (particles) are weakly independent of each other; notice that in the product states there are no correlations between the spins and each of the spins provides an independent probability distribution. 
In such cases, the asymptotic behaviors of the probability distribution can be determined by the Chernoff inequality~\cite{ref:Chernoff} (Fig.~\ref{fig:Chernoff_ineq}), which is also called as the Hoeffding inequality~\cite{ref:Hoeffding_ineq}; 
let us consider an $N$ spin system, a product state $\rho_{\Prod}=\rho_{1}\otimes\rho_{2} \otimes \cdots \otimes \rho_{N}$, and a macroscopic observables of the form $A=\sum_{i}^{N} a_{i}$ with $\|a_i\|=1$, then the probability distribution for $A$ is upperbounded by
\begin{eqnarray}
\tr (\rho_{\Prod} \Pi_{\ge x}^{A}) \le e^{-\frac{(x-\langle A \rangle)^2}{CN}} ,\label{ineq:Chernoff_bound}
\end{eqnarray}
where $\Pi_{\ge x}^{A}$ is the projector onto the eigenspace of $A$ in which the eigenvalues are larger than $x$, $\langle A \rangle:=\tr (\rho_{\Prod} A)$ and $C$ is a universal constant. 
The Chernoff inequality is originally exploited to analyze the hypothesis testing problem~\cite{ref:Chernoff,ref:Nussbaum,PhysRevLett.98.160501}. 
Here, we consider \textit{exactly independent spins}, i.e., the product states, but we expect the similar inequality also holds for \textit{weakly correlated states}: for example, 
states with the exponential clustering (the exponential decay of bi-partite correlations), ground states in non-critical regimes or with non-vanishing spectral gaps, short-range entangled states, etc. 
In these cases, the Chernoff inequality~\eqref{ineq:Chernoff_bound} is expected to hold, but has not been proved yet. 

\begin{figure}
\centering
\includegraphics[clip, scale=0.5]{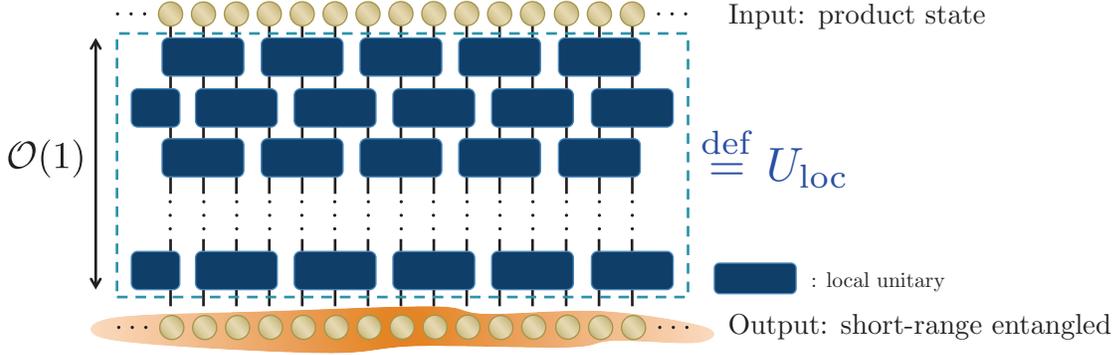}
\caption{The short-range entangled state is defined by a
constant depth unitary circuit. One layer consists of a set of unitary operations; each unitary operator
(indigo box) is applied to a small number of spins and does not overlap with each other. The depth of
the quantum circuit is bounded from above by an $\orderof{1}$ constant.
}
\label{fig:SRE}
\end{figure}

A good place to start studying this question is in the short-range entanglement class~\cite{ref:Wen-topo}, which is characterized by a constant-depth quantum circuit acting on a product state (Fig.~\ref{fig:SRE}). 
 A constant constant-depth quantum circuit is a unitary operator which can be decomposed into a product of $\orderof{1}$ unitary operators;
 here each of the unitary operators is given by the form of $U_{X_{1}}\otimes U_{X_2}\otimes \cdots\otimes U_{X_{n}}$ where $\{X_{i}\}_{i=1}^{n}$ are non-overlapping supports.  
 The short-range entangled states are often called as the ``trivial state'' as they do not show any non-local quantumness such as the macroscopic superposition~\cite{frowis2012} and the topological order~\cite{ref:Hastings10-nonZeroT}.  
We thus expect that the Chernoff inequality should hold for this class of states.
Now, the problem is equivalent to consider the probability distribution of $U_{\loc}^{\dagger} A U_{\loc}$ with respect to a product state, where $U_{\loc}$ is given by the finite-depth quantum circuit. 
Notice that the operator $U_{\loc}^{\dagger} A U_{\loc}$ is a few-body operator, because $U_{\loc}^{\dagger}  a_{i} U_{\loc}$ is still supported in the small region around the site $i$.
In the present paper, we aim to answer the following slightly general problem; for generic few-body operators, which may not be given by $U_{\loc}^{\dagger} A U_{\loc}$, can we prove the Chernoff inequality for arbitrary product states
(see Sec.~\ref{sec:settings} on the detailed definition of the few-body operators)?

We here compare the present analysis with the previous works. 
The most relevant work has been recently given by Ref.~\cite{anshu2015quantum} in a different approach, where the Chernoff inequality has been proved for one-dimensional local observables. 
There is a long history on the analysis of probability distributions in weakly or finitely correlated quantum many-body systems. 
One of the well-known examples is on the central limit theorem, which tells us that the distribution of an observable converges to a Gaussian normal distribution in thermodynamic limit, or for the infinite system size.
In Refs~\cite{ref:Hartmann,ref:vets,ref:vets2,matsui2002bosonic}, the central limit theorem has been proved for several classes of weakly correlated quantum states. 
More recently, the related Berry-Essen theorem~\cite{ref:Berry,ref:Essen} has been proved for states under the assumption of the exponential clustering~\cite{brandao2015equivalence}.  
However, in their analysis, the best convergence late is at most of $\orderof{1/\sqrt{N}}$~\cite{brandao2015equivalence} and cannot provide a good estimation for the asymptotic behavior of the probability distribution.  
The asymptotic behaviors of finite systems are often analyzed by the large deviation theorem~\cite{Touchette20091}, which has been also proved for special cases (e.g., finitely correlated spin chains~\cite{Ogata2010} and 1D high-temperature Gibbs states~\cite{Large-diviation_hiai,ref:Marco_LD}).
In other direction, for ground states with non-vanishing spectral gap, a weaker version of the Chernoff bound as $\tr (\rho \Pi_{\ge x}^{A}) \le e^{-\frac{|x-\langle A \rangle|}{\sqrt{CN}}}$ is proved for arbitrary dimensional systems~\cite{Ref:LR}, where the constant $C$ is proportional to the inverse of the spectral gap. 
So far, even in simple setups, the Chernoff-type asymptotic decay~\eqref{ineq:Chernoff_bound} has not been proved for higher or infinite dimensional systems.

\begin{figure}
\centering
\includegraphics[clip, scale=0.6]{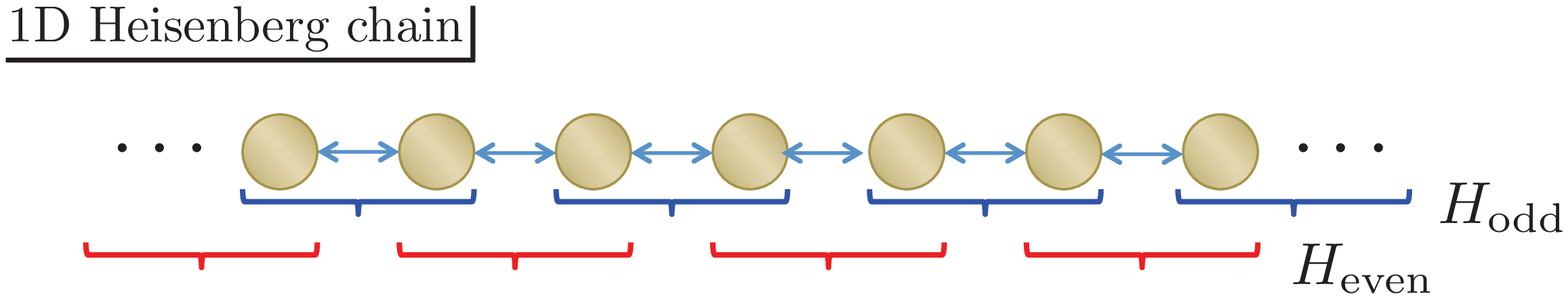}
\caption{Decomposition of operators. Let an observable be given by the Hamiltonian $H$ of the 1D Heisenberg chain. Then, by decomposing the operator into two observables, namely $H_{{\rm odd}}$ and $H_{\rm even}$. Then, each of the operators can be given by the summation of the non-overlapping local operators, and hence the Chernoff inequality~\eqref{ineq:Chernoff_bound} holds for $H_{{\rm odd}}$ and $H_{\rm even}$, respectively. However, the analysis of $H_{{\rm odd}}+H_{\rm even}$ is usually non-trivial and it is not a simple question whether the Chernoff inequality also holds for $H$.}
\label{fig:fig_decomp}
\end{figure}

In the analysis, the main difficulty lies in the fact that when the observables $A$ is a generic few-body operator the spectral analysis is usually non-trivial.
To resolve it, we aim to answer the following questions;
\textit{given probability distributions of a set of observables $\{A_{i}\}_{i=1}^{\bar{n}}$ for a quantum state $\rho$, then can we say something about the distribution function of $\sum_{i=1}^{\bar{n}}A_{i}/\bar{n}$?}
We often encounter the situation where some observables $\{A_{i}\}_{i=1}^{\bar{n}}$ are easy to analyze, while their summation $\sum_{i=1}^{\bar{n}}A_{i}/\bar{n}$ is difficult to analyze.
For example, let us consider a problem to calculate the probability distribution with respect to a product state, say $\ket{0}^{\otimes N}$ ($N$: system size), for the  Heisenberg-type observable $H=\sum_{i=1}^{N} J  (S_{i}^{x}S_{i+1}^{x} + S_{i}^{y}S_{i+1}^{y}+S_{i}^{z}S_{i+1}^{z})$; then, as shown in Fig.~\ref{fig:fig_decomp}, it is trivial to analyze each of $H_{{\rm even}}=\sum_{i={\rm even}}^{N} J  (S_{i}^{x}S_{i+1}^{x} + S_{i}^{y}S_{i+1}^{y}+S_{i}^{z}S_{i+1}^{z})$ and $H_{{\rm odd}}=\sum_{i={\rm odd}}^{N} J  (S_{i}^{x}S_{i+1}^{x} + S_{i}^{y}S_{i+1}^{y}+S_{i}^{z}S_{i+1}^{z})$ separately, 
while the analysis of $H$ is a nontrivial problem~\cite{ref:Haldane1983-HeisenbergI,ref:Haldane1983-HeisenbergII}. 
In this case, the inequality~\eqref{ineq:Chernoff_bound} trivially holds for  both of $H_{{\rm even}}$ and $H_{{\rm odd}}$.
Hence, if we can show that the asymptotic distribution behavior does not change due to the summation of $H_{{\rm even}}+H_{{\rm odd}}$,  
the probability distribution of $H$ is proved to follow the inequality~\eqref{ineq:Chernoff_bound}.
Indeed, we can generally connect the momentum generating function of $\sum_{i=1}^{\bar{n}}A_{i}/\bar{n}$ to those of $\{A_{i}\}_{i=1}^{\bar{n}}$, which plays a key role in the generalization of the Chernoff inequality.

The paper is organized as follows.
In Section~\ref{sec:settings}, we first show the fundamental setup of the system and definitions of the terms.
We also show several preliminary lemmas for the analysis.
In Section~\ref{sec:moment_func}, we first give the general statements to connect the probability distributions of different operators. 
As one application, we give an improved version of the main lemma in Ref.~\cite{arad2014connecting} which discusses the energy excitation by local perturbations.
In Section~\ref{sec:Chernoff_gene}, we show the generalized version of the Chernoff inequality~\eqref{ineq:Chernoff_bound}. 
In Section~\ref{sec:proof}, we prove all the theorems, lemmas and corollaries.
Finally, Section~\ref{sec:conc} concludes the paper.

%


%
%


\section{Notation and general setup}
\label{sec:settings}

We consider a spin system of finite volume with each spin having a $d$-dimensional Hilbert space, and we label each spin by $i=1,2,\ldots  N$. 
We denote the set of all spins by $\Lambda=\{1,2,\ldots  N\}$, a partial set of sites by $X$, and the cardinality of $X$, that is, the size of this subset, by $|X|$ (e.g. $X=\{i_1,i_2,\ldots, i_{|X|}\}$).

Also, for an arbitrary operator $A$, we define $\Pi_{\ge x}^{A}$ ($\Pi_{< x}^{A}$) as the projection operator onto the eigenspace of $A$ which are in the range $\ge x$ ($<x$).
Note that the probability distribution of $A$ for a quantum state $\rho$ having values larger than $x$ is given by $\tr (\rho \Pi_{\ge x}^{A})$.
As quantum states, we consider arbitrary $\rho$ in Sec.~\ref{sec:moment_func}, while in Sec.~\ref{sec:Chernoff_gene} we restrict $\rho$ to the product states.

Throughout the paper, we consider the few-body observables. 
To characterize their properties, we introduce two terminologies: `$q$-local' and `$g$-extensiveness.' 
 \begin{definition}[$q$-local]
 \label{def:k_local}
  We define that  $O$ is $q$-local with $q$ a positive integer if it is given by the form of
   \begin{eqnarray}
O=\sum_{|X|\le q } o_X ,
\end{eqnarray}
where each of $\{o_X\}_{X\subset \Lambda}$ is an operator supported on a finite subset of spins $X$ whose cardinality is smaller than or equal to $q$.  
Here, the subset $X$ may not be sitting next to each other on the lattice.
\end{definition}
More explicitly, a $q$-local operator can be given in the form of
 \begin{eqnarray}
\fl O=\sum_i h_i s_i + \sum_{i_1,i_2} J_{i_1,i_2} s_{i_1} s_{i_2} +\sum_{i_1,i_2,i_3} J_{i_1,i_2,i_3} s_{i_1} s_{i_2} s_{i_3} 
+\cdots +\sum_{i_1,i_2,\ldots,i_q} J_{i_1,i_2,\ldots,i_q} s_{i_1} s_{i_2} \cdots s_{i_q}, \nonumber 
\end{eqnarray}
where $\{s_i\}$ are operator bases on the $i$th spin; for example, it can be given by the Pauli matrices for $(1/2)$-spin systems, namely $\{s_i\}=\{\sigma_i^x,\sigma_i^y,\sigma_i^z\}$.

Second, we introduce the $g$-extensiveness as the normalization of operator.
 \begin{definition}[$g$-extensive]
  \label{def:g_ext}
 For a positive constant $g>0$, we say that an operator $O$ is $g$-extensive if 
  \begin{eqnarray}
  \label{def:g_extensive}
\sum_{i: i \in X} \|o_X \| \le g ,\for \forall i \in \Lambda,
\end{eqnarray}
with $\|\cdots\|$ the operator norm (i.e., the maximum singular value of operator) and $\sum_{X:X\ni {i}} $ denotes the summation with respect to the supports $X$ containing the spin $i$. 
\end{definition}
This condition implies that a local norm of one-spin is bounded by a finite constant $g$. 
A trivial algebra ensures that the norm $\|O\|$
increases at most in a linear way of $gN$ with the system size $N$: 
  \begin{eqnarray}
\|O\|=\left\| \sum_{X: X\subset \Lambda}  o_X \right\| \le \sum_{X: X\subset \Lambda} \| o_X\| \le \sum_{i=1}^{N} \sum_{X: X\ni i} \|o_X\| \le  \sum_{i=1}^{N} g =gN,
\end{eqnarray}
where we note that the $q$-locality of $O$ is not assumed.

For arbitrary quantum state $\rho$ and operator $A$,
we define the moment generating function $e^{M(\rho, A, \tau)}$ as follows: 
\begin{eqnarray}
M(\rho, A, \tau) := \log \left[ \tr (e^{\tau A} \rho) \right].
\end{eqnarray}
Notice that $M(\rho, A, \tau)$ contains information on the asymptotic behavior of the probability distribution. 
For example, let us consider a case where $M(\rho, A, \tau)$ is bounded from above by
\begin{eqnarray}
M(\rho, A, \tau) \le  c_{1}  \tau^{2} + c_{2}
\end{eqnarray}
with $\tr (\rho A)=0$,
where $c_{1}$ and $c_{2}$ are positive constants.
We then have 
$
e^{\tau x}  \tr (\Pi_{\ge x}^{A} \rho) \le  \tr (e^{\tau A} \rho) \le e^{c_{1} \tau^{2} + c_{2}}
$ 
for $x\ge 0$, 
which yields 
\begin{eqnarray}
\label{moment_gauss_ineq}
\tr (\Pi_{\ge x}^{A} \rho) \le   \exp \left(-\frac{x^{2}}{4 c_{1}}  + c_{2}\right)
\end{eqnarray}
by choosing $\tau = x/(2c_{1})$. The same inequality holds for $\tr (\Pi_{\le - x}^{A} \rho) $.

\subsection{Preliminaries}
We here show three basic lemmas. 
First, for the analysis of generic few-body observables, we often need to treat the multi-commutators.
We give two lemmas as useful technical tools in the analysis.  
Second, we formulate the Chernoff inequality for operators which are given by summation of independent local observables. 

For the norm of multi-commutators, we can prove the following lemma (see Lemma~3 in Ref.~\cite{Kuwahara201696}):
\begin{lemma} \label{norm_multi_commutator1}
Let $\{A_i\}_{i=1}^n$ be $k_i$-local and $g_{i}$-extensive, respectively, and 
$O_X$ be an arbitrary operator supported in a subset $X$.
Then, the norm of the multi-commutator 
$ [A_n,[A_{n-1}, [\cdots, [A_1,O_X ]\cdots]]$ 
is bounded from above by
\begin{eqnarray}
\| [A_n,[A_{n-1}, [\cdots, [A_1,O_X ]\cdots]]\| \le \prod_{m=1}^n (2  g_{m} K_m) \|O_X\|, \label{fundamental_ineq}
\end{eqnarray}
where $K_m := |X|+ \sum_{i\le m-1} k_i $.
\end{lemma}

\begin{figure}
\centering
\includegraphics[clip, scale=0.5]{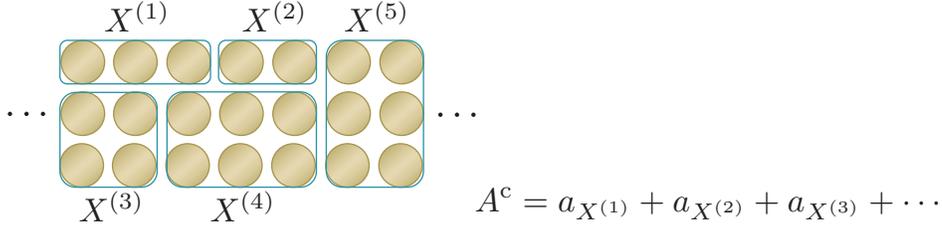}
\caption{Decomposed operators in Lemma~\ref{decomposition}. The lemma says that we can decompose an arbitrary few-body operator into a set of simple operators $\{A_{j}^\co\}_{j=1}^{\bar{n}}$, each of which is given by the summation of non-overlapping local operators. 
The terminology ``non-overlapping" means that the local components in $A_j^\co$, namely $\{a_{X_j^m}\}_{m=1}^{N_j}$ in Eq.~\eqref{construction_Commuting_Hm}, are supported in subsets $\{X_j^m\}_{m=1}^{N_j}$ ($|X_j^m|\le k$) which do not overlap with each other. 
}
\label{fig:fig_commuting}
\end{figure}

Then, we consider a problem to decompose a few-body operator $A$ into a summation of operators $\{A_{i}\}_{i}^{\bar{n}}$, each of which is easy to analyze. For example, in considering a spatially-local Hamiltonian on a finite-dimensional lattice, the similar decomposition to Fig.~\ref{fig:fig_decomp} is always possible by taking $\bar{n}=\orderof{D}$ with $D$ the system dimension. 
On the other hand, for generic few-body observables $A$~(e.g., a local operator on infinite-dimensional graph), existence of such a decomposition is non-trivial.  
Actually, we can also ensure the existence of the following decomposition for general $k$-local and $g$-extensive operators~\cite{kuwahara2015locality}:
\begin{lemma} \label{decomposition} 
Let $A$ be an arbitrary $k$-local and $g$-extensive operator. 
Then, we can always find a decomposition of $A$ into a summation of $\bar{n}$ operators $\{A_{j}^\co\}_{j=1}^{\bar{n}}$ such that (Fig.~\ref{fig:fig_commuting})
\begin{eqnarray}
\left \|A-\frac{1}{\bar{n}} \sum_{j=1}^{\bar{n}} A_j^\co \right \| =  \orderof{N/\bar{n}}, 
\label{SC_decomp}
\end{eqnarray}  
with
\begin{eqnarray}
\fl&A_j^{\co} = a_{X_j^{(1)}} + a_{X_j^{(2)}} + \cdots + a_{X_j^{(N_j)}} ,      \nonumber \\
\fl&s.t.\quad  \| a_{X_j^{(m)}}\| \le gk \quad {\rm and} \quad  X_j^{(m)} \cap X_j^{(m')} =\emptyset \quad {\rm for}\quad  \forall m,m' \in \{1,2,\ldots,N_j\},  \label{construction_Commuting_Hm}
\end{eqnarray}
where each of $\bigl\{ a_{X_j^{(1)}}, a_{X_j^{(2)}}, \ldots, a_{X_j^{(N_{j})}}\bigr \}$  is proportional to a local component of $A$ and $N_j \le N/k$ for $j=1,2,\ldots,N$. 
Note that $\{ A_j^\co \}_{j=1}^{\bar{n}}$ are now $k$-local and $(gk)$-extensive. 
With the integer $\bar{n}$ infinitely large, the operator $A$ can be decomposed into summation of operators of the form~\eqref{construction_Commuting_Hm}.
\end{lemma}

We formalize the quantum Chernoff inequality as the upper bound of the moment generating function $M(A,\rho,\tau)$:

\begin{lemma}[The Chernoff inequality for product states] \label{Chernoff:def} 
Let $A$ be a summation of non-overlapping local operators:
\begin{eqnarray}
\fl &A = \sum_{i=1}^{N_A} a_{X_i}  \nonumber \\
\fl &  s.t. \quad   \|a_{X_i}\| \le 1,\quad |X_{i}|\le k \quad {\rm and} \quad X _i \cap X_j = \emptyset \quad \forall i,j \in \{1,2,\ldots,N_A \} ,\label{assump:Chernoff_ineq_operator}
\end{eqnarray}
and $\rho_{\Prod}$ be a product state in the form of $\rho_1 \otimes \rho_2 \otimes \cdots \otimes \rho_N$. 
For the simplicity, we set $\tr (\rho A) =0$.  
Then, the Chernoff inequality gives the Gaussian decay of the probability distribution of $\rho_{\Prod}$ with respect to $A$, or equivalently 
\begin{eqnarray}
M(A,\rho_{\Prod},\tau) \le C \frac{N}{k} \tau^2 \label{Chernoff_ineq_moment}
\end{eqnarray}
with $C$ a universal constant. 
\end{lemma}
From the lemma, we immediately obtain
\begin{eqnarray}
\tr \left( \Pi^A_{\ge  x}\rho_{\Prod}   \right) \le  \exp\left( -\frac{k x^2}{4C N}\right)  \label{Prob_gaussian_Cher1}
\end{eqnarray}
by following the same discussion as the derivation of \eqref{moment_gauss_ineq}.


\section{Upper bound on the moment generating function} \label{sec:moment_func}
In this section, we consider the following problem.
Let $\rho$ be an arbitrary quantum state and $\{A_{j}\}_{j=1}^{\bar{n}}$ be a set of arbitrary operators. 
We also assume that we know the probability distribution of each of $\{A_{j}\}_{j=1}^{\bar{n}}$. 
Here, we consider how to connect the probability distributions of $\{A_{j}\}_{j=1}^{\bar{n}}$ to that of their summation $\sum_{j=1}^{\bar{n}}A_{j}/\bar{n}$. 

We begin with the most general statement without any assumption on the operator $\{A_{j}\}_{j=1}^{\bar{n}}$. 
For this question, we can trivially obtain information on the average and the standard variance. 
For the simplicity, we discuss the case of $\bar{n}=2$ and consider two observables $A_{1}$ and $A_{2}$.  
We denote the average and the variance by $\mu$ and $\sigma$, respectively, and set the averages to be zero ($\mu_{1}=\mu_{2}=0$). 
Then, for the summation $A_{1}+A_{2}$, we have
\begin{eqnarray}
\tr [\rho (A_1+A_2)] = \tr (\rho A_1)+\tr (\rho A_2)=\mu_1+\mu_2=0 
\end{eqnarray}
and
\begin{eqnarray}
\sqrt{\tr [\rho (A_1+A_2)^2]} &= \sqrt{\tr (\rho A_1^2)+\tr (\rho A_2^2)+\tr (\rho A_2A_1)+\tr (\rho A_1A_2)} \nonumber \\
&\le  \sqrt{\left( \sqrt{\tr (\rho A_1^2)} +\sqrt{\tr (\rho A_2^2)} \right)^2}=\sigma_1+\sigma_2,
\end{eqnarray}
where we apply the Schwartz inequality $\tr (\rho A_2A_1) \le \sqrt{\tr (\rho A_1^2) \tr (\rho A_2^2)}$. 
By using the Chebyshev inequality, we can ensure that the probability distribution for $A_{1}+A_{2}$ decays faster than
\begin{eqnarray}
\tr \left(\rho \Pi_{\ge x}^{A_{1}+A_{2}}\right) \le  \left( \frac{\sigma_{1}+\sigma_{2}}{|x|} \right)^{2} ,\label{ineq:Cheby_bound}
\end{eqnarray}
where $\Pi_{\ge x}^{A_{1}+A_{2}}$ is the projection operator onto the eigenspace of $A_{1}+A_{2}$ which are in $[x,\infty)$.

From the bound~\eqref{ineq:Cheby_bound}, we cannot know information on over third order moments like $\tr [\rho (A_1+A_2)^k] $ with $k\ge 3$.   
We aim to obtain much better bound by considering additional assumptions to the observables $A_{1}$ and $A_{2}$. 
In more details, we restrict the observables to be $k$-local and $g$-extensive. 
This condition is one of the most natural assumptions when we analyze realistic quantum many-body systems. 
Nevertheless, the recent studies have shown that this ``few-body'' assumption gives us a fruitful information on the fundamental properties. 
For arbitrary sets of few-body observables $\{A_{i}\}_{i=1}^{\bar{n}}$, we obtain strong upper bounds on the moment generating function. We start from the case of $\bar{n}=2$:

\begin{theorem} \label{norm_momentum_generator}
Let $A$ and $B$ be $k$-local $g$-extensive operators, respectively.
Then, for arbitrary quantum state $\rho$ and $|\tau| < 1/(4\lambda)$, we have 
\begin{eqnarray}
M (A+B,\rho,\tau) \le  \frac{M (2A,\rho,\tau)+M (2B,\rho,\tau)}{2}+\frac{3\tau^2}{2} \|[A,B]\| ,
 \label{norm_momentum_generator_thm}
\end{eqnarray}
where $\lambda$ is defined by
\begin{eqnarray}
\lambda := 2gk.
\label{def:lambda}
\end{eqnarray}
\end{theorem}
By using Lemma~\ref{norm_multi_commutator1} and the $g$-extensiveness of operators, the commutator norm $\|[A,B]\| $ is bounded from above by $\|[A,B]\|\le \lambda gN$. Hence, we can generally obtain the inequality 
\begin{eqnarray}
M (A+B,\rho,\tau) \le  \frac{M (2A,\rho,\tau)+M (2B,\rho,\tau)}{2}+ \frac{3\lambda g N \tau^2}{2} .
 \label{Schwartz_A_B_est2}
\end{eqnarray}
Here, we would like to emphasize that we do not need \textit{any assumption} on the quantum state $\rho$ like the product states or the pure states.

We can easily extend the results to general summations: 
\begin{corol} \label{norm_momentum_generator2}
Let $\{A_j\}_{j=1}^{\bar{n}}$ be arbitrary $k$-local $g$-extensive operators, respectively, and $A$ be defined by their summation:
\begin{eqnarray}
A= \frac{1}{\bar{n}} \sum_{j=1}^{\bar{n}} A_j .
\end{eqnarray}
Then, for arbitrary quantum state $\rho$ and $|\tau| < 1/(4\lambda)$, we have 
\begin{eqnarray}
M (A ,\rho,\tau) \le  \frac{1}{2^{m_0}} \sum_{j=1}^{\bar{n}} M \left ( \frac{2^{m_0}}{\bar{n}} A_j ,\rho,\tau \right) + \frac{3\lambda g N \tau^2}{8}  m_0 ,
 \label{norm_momentum_generator_thm2}
\end{eqnarray}
where $m_0= \lceil \log_2 n \rceil$. 
\end{corol}
The error term is proportional to $N \tau^2 \log \bar{n}$, and hence as long as we consider a summation of finite number of operators ($\bar{n}=\orderof{1}$), 
the term is not influential.
This $\log \bar{n}$ dependence becomes dominant in considering the Chernoff inequality for general few-body operators (See Sec.~\ref{Chernoff_general_dim}), where $\bar{n}$ is at least as large as ${\rm Poly} (N)$. 
This theorem plays central roles in deriving the Chernoff inequality~\eqref{ineq:Chernoff_bound} for general few-body operators. 


We further show that much stronger inequality holds if we add the assumption that the distributions are completely localized for $\rho$:
 \begin{theorem} \label{norm_momentum_generator_localize}
Let  $A$ be defined by
\begin{eqnarray}
A= \frac{1}{\bar{n}}\sum_{j=1}^{\bar{n}} A_j^{\co} ,
\end{eqnarray}
where each of $\{A_j^{\co}\}_{j=1}^n$ is $k$-local, $g$-extensive and is given by a summation of non-overlapping local operators (see Fig.~\ref{fig:fig_commuting}).
We also assume that for a quantum state $\rho$ the distribution of $\{A_j^{\co}\}_{j=1}^n$ is exactly localized with a width $2\sigma$ and $\tr (\rho A_{j})=0$ for $j=1,2,\ldots,n$; that is, we assume
\begin{eqnarray}
\label{a:assump_localize}
\rho \Pi_{> \sigma}^{A_{j}} =\rho \Pi_{<- \sigma}^{A_{j}} =0
\end{eqnarray}
for $j=1,2,\ldots,n$.
Then, the moment generating function $M (A,\rho,\tau)$ is upperbounded by
\begin{eqnarray}
M (A,\rho,\tau) \le \frac{-\sigma}{\lambda} \log (1-\lambda |\tau|),
\end{eqnarray}
where $\lambda$ is defined in Eq.~\eqref{def:lambda}.
\end{theorem}
This implies that the probability distribution of $A$ is also localized in a finite range with exponentially decaying errors; it is ensured by
\begin{eqnarray}
\tr \left (\Pi_{\ge x}^{A} \rho  \right) &\le  e^{-\tau x}  \tr \left (e^{\tau A}\rho  \right)\le e^{-\tau x}  (1-\lambda |\tau|)^{-\sigma/\lambda}  ,
\end{eqnarray}
which yields,
\begin{eqnarray}
\tr \left (\Pi_{\ge x}^{A} \rho  \right) &\le  \left(\frac{x}{\sigma} \right)^{\sigma/\lambda}e^{-(x-\sigma)/\lambda} \for x\ge \sigma,
\label{norm_momentum_generator2_localize}
\end{eqnarray}
where we choose $\tau=(x-\sigma)/(\lambda x)$. We can obtain the same inequality for $\tr \left (\Pi_{\le -x}^{H} \rho  \right)$.

As one application of Theorem~\ref{norm_momentum_generator_localize}, we show the improvement of the following inequality~\cite{arad2014connecting}; 
let $H$ be a Hamiltonian which is $k$-local and $g$-extensive and $A$ be an arbitrary $q$-local operator. Then, the 
energy excitation due to the operator $A$ is exponentially suppressed as 
\begin{eqnarray}
\| \Pi^{H}_{\ge x} A \ket{\Omega} \| \le \|A\| \exp\left(- \frac{x}{5k\lambda} +q\right) ,
\label{AKL:Basic}
\end{eqnarray}
where $\ket{\Omega}$ is the ground state of the Hamiltonian $H$ as $H\ket{\Omega}=0$.
This kind of inequality has been first introduced to analyze an effective Hamiltonian which governs low-energy regimes~\cite{arad2014connecting}. 
It plays central roles in connecting spectral gap and the fundamental properties of ground states~\cite{Ref:LR,ref:Arad-AL13,arad2016rigorous}.

On the other hand,  due to the coefficient $\|A\|$ in the right-hand side, arbitrary large energy can be still locally excited \textit{with very low-probability.} 
To show the point, we consider the case where the ground state $\ket{\Omega}$ is decomposed as follows:
\begin{eqnarray}
\ket{\Omega} = \lambda_{0}\ket{0}\otimes \ket{\phi_{0}} + \lambda_{1} \ket{1} \otimes \ket{\phi_{1}},
\end{eqnarray}
where $|\lambda_{0}|^{2}+|\lambda_{1}|^{2}=1$.
We now choose $A=\ket{0}\bra{0}/\lambda_{0}$ or $A=\ket{1}\bra{1}/\lambda_{1}$, and then the inequality~\eqref{AKL:Basic} reads
\begin{eqnarray}
\fl \| \Pi^{H}_{\ge x}\ket{0}\otimes \ket{\phi_{0}} \| \le   \frac{1}{\lambda_{0}}\exp\left(- \frac{x}{5k\lambda} +1\right),\quad \| \Pi^{H}_{\ge x}\ket{1}\otimes \ket{\phi_{1}} \| \le   \frac{1}{\lambda_{1}}\exp\left(- \frac{x}{5k\lambda} +1\right),
\label{energy:excitation:locally}
\end{eqnarray}
where $A$ is 1-local ($q=1$) and $\|A\|=1/\lambda_{0}$ ($1/\lambda_{1}$).
Thus, if $\lambda_{0}$ or $\lambda_{1}$ is as small as $e^{-\orderof{N}}$, the macroscopic energy of $\orderof{N}$ can be locally excited with the probability of $e^{-\orderof{N}}$. It seems rather strange, and hence
we expect that the inequality can be improved in the following way:
\begin{eqnarray}
\| \Pi^{H}_{\ge x} A \ket{\Omega} \| \le \|A\ket{\Omega}\| \exp\left(- \frac{x}{5k\lambda} +q\right) ,
\end{eqnarray}
which modifies the inequality~\eqref{energy:excitation:locally} as  $\| \Pi^{H}_{\ge x}\ket{0}\otimes \ket{\phi_{0}} \| \le e^{- \frac{x}{5k\lambda} +1}$ and $\| \Pi^{H}_{\ge x}\ket{1}\otimes \ket{\phi_{1}} \| \le e^{- \frac{x}{5k\lambda} +1}$, where the high-energy excitation is impossible even probabilistically.

We can indeed improve the inequality~\eqref{AKL:Basic} in the case where the Hamiltonian is frustration free.
A frustration-free Hamiltonian satisfies the following property: the Hamiltonian can be written as a sum of terms such
that the lowest energy states of the full Hamiltonian are also the lowest energy states of each individual term. In other words,
the global ground states are also local ground states.
Then, we can prove the following corollary:

\begin{corol} \label{norm_momentum_generator_appli}
Let $H$ be a frustration free $k$-local Hamiltonian with $g$-extensiveness, i.e., $h_{X} \ket{\Omega}=0$ for $\forall h_{X}$ with 
$H=\sum_{|X|\le k}h_{X}$. We also denote the ground state of $H$ by $\ket{\Omega}$  ($H\ket{\Omega}=0$).
Then, for any $q$-local operator $A$, we have
\begin{eqnarray}
\| \Pi^{H}_{\ge x} A \ket{\Omega}\| \le \|A \ket{\Omega}\|   \left(\frac{x}{\lambda q} \right)^{q/(2k)} \exp \left( - \frac{x}{2 k \lambda} +\frac{q}{2 k} \right) \for x\ge \lambda q . 
\label{general_ver_AKL}
\end{eqnarray}
\end{corol}

\section{The Chernoff inequality} \label{sec:Chernoff_gene}

\begin{figure}
\centering
\subfigure[Operators on finite dimensional lattices]{
\includegraphics[clip, scale=0.55]{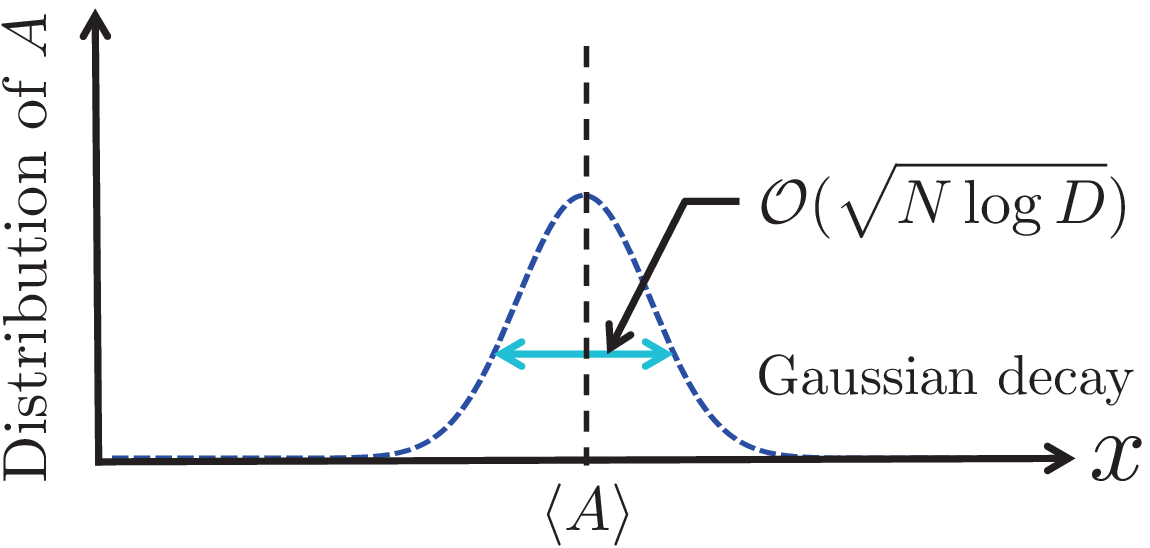}
}
\subfigure[Generic few-body operators]{
\includegraphics[clip, scale=0.55]{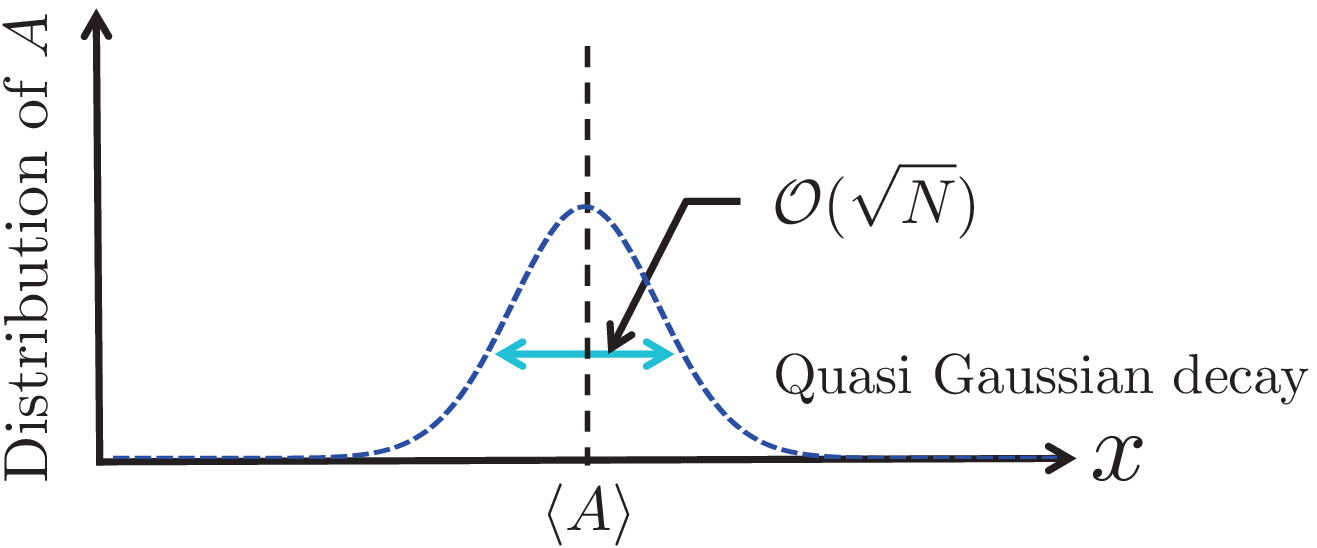}
}
\caption{Generalization of the Chernoff inequality. When we consider spatially local observables on finite dimensional lattices, the operators can be decomposed into $\orderof{D}$ operators each of which is given by the summation of non-overlapping local operators. In this cases, the probability distribution of $A$ for a product state $\rho_{\Prod}$ decays faster than the Gaussian decay, but with the variance of $\orderof{N \log D}$ (Lemma~\ref{Chernoff_finite_dim}). On the other hand, for generic few-body operators or local operators on infinite dimensional lattices, we obtain slightly weaker decays than the strict Gaussian, $e^{-\orderof{x^2/[N\log (x /\sqrt{N})]}}$ (Corollary~\ref{Chernoff_infinite_dim2}). 
}
\label{fig:Chernoff_theory}
\end{figure}

In this section, we extend the Chernoff inequality~\eqref{Chernoff_ineq_moment} to more general cases than the standard ones~\eqref{assump:Chernoff_ineq_operator}, which is restricted to the summation of non-overlapping local operators. 
We here consider the probability distribution of generic few-body observables for a product state $\rho_{\Prod}$. 
We summarize the present results in Fig.~\ref{fig:Chernoff_theory}.

\subsection{Finite dimensional systems}
In considering finite-dimensional systems, we have to define the structure of the system explicitly (e.g., the square lattice).
For the simplicity, we here restrict ourselves to $D$-dimensional regular lattices. 
On this lattice system, we define the distance $\dist(X, Y)$ as the shortest-path length which one needs to connect the two partial sets $X$ and $Y$.
When we consider ``spatially local operators," we introduce the following additional assumption to the operator 
 \begin{eqnarray}
A=\sum_{|X|\le k}a_{X} , \quad s.t. \quad \sum_{X: X \ni i,  \diam (X) \ge r } \| a_X \| =0 \for \forall i \in \Lambda , \label{finite_range_interaction_def}
\end{eqnarray}
where $r$ is an $\orderof{1}$ constant and $\diam (X):= \sup_{\{i,j\} \in X} \dist(i,j)$.

First, we consider the case that the operator $A$ is given by operators which satisfy \eqref{finite_range_interaction_def}. 
This assumption ensures that we can always decompose the operator $A$ into
\begin{eqnarray}
A=\frac{1}{\bar{n}} \sum_{j=1}^{\bar{n}} A_j^\co \quad {\rm with} \quad \bar{n} =\orderof{D} , \label{SC_decomp_finite_dim}
\end{eqnarray}  
where each of $\{A_j^\co\}_{j=1}^{\bar{n}}$ is given by the summation of non-overlapping local operators. 
For example, let us consider an operator $A_{1D}$ on a one-dimensional lattice with nearest-neighbor couplings ($r=1$ in Eq.~\eqref{finite_range_interaction_def}):
 \begin{eqnarray}
A_{1D}= \sum_{i=1}^{N-1} a_{i,i+1} ,
\end{eqnarray}
where $\{ a_{i,i+1}\}_{i=1}^{N-1}$ are arbitrary operators defined on the sites $\{i,i+1\}_{i=1}^{N-1}$, respectively, and satisfy $\|a_{i,i+1}\|\le g$ with $g$ a constant.
This can be decomposed as $A_{1D}= (A_{1}^{\co}+A_{2}^{\co})/2$ with 
 \begin{eqnarray}
A_{1}^{\co} = \sum_{i={\rm even}}^{N-1} 2 a_{i,i+1} ,\quad A_{2}^{\co} = \sum_{i={\rm odd}}^{N-1}  2a_{i,i+1}.
\end{eqnarray}
The operators $A_{1}^{\co}$ and $A_{2}^{\co}$ are now $2$-local and $(2g)$-extensive, respectively. 
Notice that the Chernoff inequality~\eqref{Chernoff_ineq_moment} holds for $A_{1}^{\co}$ and $A_{2}^{\co}$, respectively.

Here, from Corollary~\ref{norm_momentum_generator2}, we give the following generalization of the Chernoff inequality in Lemma~\ref{Chernoff:def}:
\begin{lemma} \label{Chernoff_finite_dim}
(Chernoff inequality for finite dimensional systems) \\
Let $\rho_{\Prod}$ be a product state $\rho_{\Prod}$ and $A$ be an operator which is given by Eq.~\eqref{SC_decomp_finite_dim} with each of $\{ A_j^\co\}_{j=1}^{\bar{n}}$ $k$-local and $g$-extensive.
We also set $\tr (\rho_{\Prod} A_j^\co) =0$ for $j=1,2,\ldots,\bar{n}$.
Then, the upper bound of the moment generating function for $\rho_{\Prod}$ is given by
\begin{eqnarray}
M (A,\rho_\Prod,\tau) \le \tilde{C} N\tau^2 
\end{eqnarray}
for $|\tau| \le 1/(4\lambda)$, where 
\begin{eqnarray}
\tilde{C} :=\frac{2g^{2} C}{k} +\frac{3\lambda g}{2} \lceil \log_2 \bar{n}\rceil.   \label{def_of_tilde_C_tau}
\end{eqnarray}
Now, the parameter $\tilde{C}$ increases logarithmically with the system dimension $D$.
\end{lemma}

By generalizing the approach in Ref.~\cite{anshu2015quantum} on 1D systems, we can also obtain a similar bound but a weaker estimation of $\tilde{C}=e^{\orderof{D}}$~\footnote{A. Anshu, private communication.}.

We can immediately apply this lemma to the probability distribution.
We have an inequality 
\begin{eqnarray}
\tr \left( \Pi^A_{\ge x}\rho_{\Prod}   \right) \le e^{-\tau x+ \tilde{C} N \tau^2 }  \label{Prob_gaussian_Cher_finite_dim}
\end{eqnarray}
for $x\ge 0$ and $\tau \le 1/(4\lambda)$. By choosing the parameter $\tau$ appropriately, the inequality~\eqref{Prob_gaussian_Cher_finite_dim} reduces to
\begin{eqnarray}
\label{ineq:distribution_gaussian_chernoff}
\tr \left( \Pi^A_{\ge  x}\rho_{\Prod}   \right) \le \max \left( e^{-\frac{x^2}{4\tilde{C} N}} , e^{-\frac{x}{8\lambda}} \right) .
\end{eqnarray} 
The same inequality holds for $\tr \left( \Pi^A_{\le -x}\rho_{\Prod}   \right)$.

\subsection{Infinite dimensional systems (Generic few-body operators)} \label{Chernoff_general_dim}
In the case where the operator $A$ is given by a general $k$-local and $g$-extensive operator, 
the number of decomposed operator, namely $\bar{n}$, can be arbitrarily large as in Lemma~\ref{decomposition}. 
Thus, we can no longer expect that $\log_2 \bar{n} $ is an $\orderof{1}$ constant; $\tilde{C}(\tau)$ in \eqref{def_of_tilde_C_tau} is infinitely large in the limit of $\bar{n}\to \infty$. We can still prove the following weaker statement for arbitrarily large $\bar{n}$.

\begin{theorem} \label{Chernoff_infinite_dim}
Let $A$ be given in the form of Eq.~\eqref{SC_decomp_finite_dim} with each of $\{ A_j^\co\}_{j=1}^{\bar{n}}$ $k$-local and $g$-extensive, where the number $\bar{n}$ may be infinitely large.
Then, the probability distribution of $A$ for $\rho_{\Prod}$ is bounded from above by
\begin{eqnarray} 
\tr \left( \Pi^A_{\ge x}\rho_{\Prod}   \right) \le c_1 \exp \left( - c_0 \frac{x^{2}}{N \log (x/\sqrt{N})}  \right) 
\label{ineq:Chernoff_infinite_dim}
\end{eqnarray}
with $x>0$, where we set $\tr (A_j^\co \rho_{\Prod})=0$ for $j=1,2,\ldots, \bar{n}$ and $\{c_0,c_1\}$ are positive constants of $\orderof{1}$ which only depend on $k$, $g$ and $C$.
The same upper bound is also given for $\tr \left( \Pi^A_{\le - x} \rho_{\Prod} \right)$. 
\end{theorem}

This inequality means that the probability distribution asymptotically decays as $e^{-\orderof{N/\log N}}$ for $x=\orderof{N}$. 
This decay is looser than \eqref{ineq:distribution_gaussian_chernoff} for $\bar{n} = \orderof{1}$, whereas 
even for $\bar{n} \to \infty$ the inequality~\eqref{ineq:Chernoff_infinite_dim} gives us the meaningful bound. 
From Lemma~\ref{decomposition}, we immediately generalize the Chernoff inequality to generic few-body observables.  
 
\begin{corol} \label{Chernoff_infinite_dim2} 
(Chernoff bound for generic few-body operators)  \\
For arbitrary $k$-local and $g$-extensive operator $A$ and product state $\rho_{\Prod}$ with $\tr(\rho_{\Prod} A)=0$, 
we have the upper bound of
\begin{eqnarray} 
\tr \left( \Pi^A_{\ge x}\rho_{\Prod}   \right) \le c'_1 \exp \left( - c'_0 \frac{x^{2}}{N \log (x/\sqrt{N})}  \right) 
\label{ineq:Chernoff_infinite_dim_generic_few}
\end{eqnarray}
for $x>0$, where $c'_0$ and $c'_1$ are positive constants of $\orderof{1}$ which only depend on $k$, $g$ and $C$. 
We have the same inequality for $\tr \left( \Pi^A_{\le - x} \rho_{\Prod} \right)$. 
\end{corol}


\section{Proof of results} \label{sec:proof}

\subsection{Proof of Theorem~\ref{norm_momentum_generator}}
\label{sec:expE}

We begin with the following inequality:
\begin{eqnarray}
\fl \tr ( e^{\tau(A+B)}\rho) &= \sum_{j=1} p_j\bra{\psi_j}e^{\tau(A+B)}\ket{\psi_j} =  \sum_{j=1} p_j \bra{\psi_j} e^{\tau B} e^{-\tau B} e^{\tau(A+B)}e^{-\tau A}e^{\tau A} \ket{\psi_j} \nonumber \\
\fl&\le \|e^{-\tau B} e^{\tau(A+B)}e^{-\tau A}\|   \sum_{j=1} p_j \| e^{\tau A} \ket{\psi_j}\|  \cdot \|e^{\tau B} \ket{\psi_j}\|\nonumber \\
\fl&= \|e^{-\tau B} e^{\tau(A+B)}e^{-\tau A}\|   \sum_{j=1} \sqrt{p_j \bra{\psi_j} e^{2\tau A} \ket{\psi_j}} \sqrt{p_j \bra{\psi_j} e^{2\tau B} \ket{\psi_j}} \nonumber \\
\fl&\le \|e^{-\tau B} e^{\tau(A+B)}e^{-\tau A}\|\sqrt{ \tr ( e^{2\tau A}\rho)} \sqrt{  \tr ( e^{2\tau B}\rho)  } ,\label{Schwartz_A_B_est}
\end{eqnarray}
where we define $\rho= \sum_{j} p_j \ket{\psi_j}\bra{\psi_j}$ and used the Schwartz inequality in the second inequality.
Then, our task is to estimate the norm of $\|e^{-\tau B} e^{\tau(A+B)}e^{-\tau A}\|$.
Here, the conditions for the operators, namely $k$-locality and $g$-extensiveness, play central roles in the estimation.  

For this purpose, we define by $e^{-x B} e^{x(A+B)}e^{-x A} =: U(x)$, and obtain 
\begin{eqnarray}
\frac{dU(x)}{dx} &= - B U - U  A + e^{-x B} e^{x(A+B)} (A+B)  e^{-x A} \nonumber \\
&=-  B U  +  e^{-x B}  e^{x (A+B)}    B e^{-x (A+B)}e^{x B} U=  [\tilde{B}(x) -B) ]U,
\end{eqnarray}
where $\tilde{B}(x) := e^{-x B}  e^{x(A+B)}B e^{-x(A+B)}e^{x B}$.
We thus obtain 
\begin{eqnarray}
U(\tau) &= \mathcal{T} \left [e^{\int_0^{\tau} [\tilde{B}(x) -B) ] dx } \right] \label{U_tau_express}
\end{eqnarray}
with $\mathcal{T} [\cdot]$ the time-ordering operator.
We then arrive at the inequality of
\begin{eqnarray}
\| U(\tau) \|&\le  e^{\int_0^{|\tau|} \| \tilde{B}(x) -B \| dx }.
\end{eqnarray}
We notice that $\tilde{B}(x)$ is given by
\begin{eqnarray}
\fl \tilde{B}(x)-B =\sum_{m,n=0}^\infty \frac{x^{n+m}}{n!m!}\ad_B^n\bigl[\ad_{A+B}^m (B)\bigr]-B
 =\sum_{n=0}^\infty  \sum_{m=1}^\infty \frac{x^{n+m}}{n!m!}\ad_B^n \left[\ad_{A+B}^m (B) \right],
\end{eqnarray}
where $\ad_B^n(\cdots)=[B, \ad_B^{n-1}(\cdots)]$.

By using the basic lemma~\ref{norm_multi_commutator1}, we obtain
\begin{eqnarray}
\| \ad_{A+B}^m (B) \| \le \| \ad_{A+B}^{m-1} ([A,B]) \| \le  (2\lambda)^{m-1}m! \|[A,B]\|, 
\end{eqnarray}
where we defined $\lambda:=2gk$ and use the fact that $A+B$ is at most $(2g)$-extensive.
Similarly, we have
\begin{eqnarray}
\| \ad_B^n \left[  \ad_{A+B}^m (B) \right]\| \le  2^{m-1} \lambda^{n+m-1} (n+m)! \|[A,B]\| .
\end{eqnarray}
We therefore calculate the upper bound of $\| \tilde{B}(x) -B\|$ as
\begin{eqnarray}
\| \tilde{B}(x) -B\|&\le  \sum_{n=0}^\infty  \sum_{m=1}^\infty \frac{|x|^{n+m}}{n!m!} 2^{m-1} \lambda^{n+m-1} (n+m)! \|[A,B]\| \nonumber \\
&=\sum_{m=1}^\infty \frac{(\lambda |x|)^{m}}{2\lambda} 2^{m} (1-\lambda |x|)^{-m-1} \|[A,B]\| \nonumber \\
&=  \frac{|x|}{(1-3\lambda|x|)(1-\lambda |x|)} \|[A,B]\|
\end{eqnarray}
for $\lambda |x| <1/3$, where we use the taylor expansion of $(1-x)^{-m-1}= \sum_{n=0}^{\infty}\frac{x^{n}}{n!} (m+1) (m+2) \cdots (m+n)$ in the first equality. 
Because we have assumed $|\tau |\lambda < 1/4$ for $\tau$, we finally obtain
\begin{eqnarray}
\| U(\tau) \|&\le   \exp \left[ \frac{\|[A,B]\|}{\lambda^2} \int_0^{\lambda |\tau|} \frac{t}{(1-t)(1-3t)} dt   \right] \nonumber \\
 &=\exp \left[ \frac{\|[A,B]\|}{\lambda^2} (2^{-1} \log (1-\lambda|\tau|) -  6^{-1} \log (1-3\lambda|\tau|) )   \right]  \nonumber \\
&\le \exp \left[ \frac{ 3\tau^2}{2} \|[A,B]\|  \right] ,\label{U_2_tau_norm_bound}
\end{eqnarray}
where we use the inequality $2^{-1} \log (1-x) -  6^{-1} \log (1-3x) \le 3x^{2}/2$ for $0\le x\le1/4$.

Thus, from the inequalities~\eqref{Schwartz_A_B_est} and \eqref{U_2_tau_norm_bound}, we arrive at the inequality~\eqref{norm_momentum_generator_thm} 
for $|\tau| < 1/(4\lambda)$. 
This completes the proof. $\square$

\subsection{Proof of Corollary~\ref{norm_momentum_generator2}}
Because of $\bar{n} \le 2^{m_{0}}$, we first denote 
\begin{eqnarray}
A=\frac{1}{\bar{n}}\sum_{j=1}^{\bar{n}} A_j =\sum_{j=1}^{2^{m_0}} \tilde{A}_j,  
\end{eqnarray}
where $\tilde{A}_{j} = A_{j}/\bar{n}$ for $ j \le \bar{n}$ and $\tilde{A}_j=0$ for $n< j \le 2^{m_0}$.
Note that each of $\{\tilde{A}_{j}\}_{j=1}^{2^{m_{0}}}$ is now $k$-local and $(g/\bar{n})$-extensive.
For the proof, we begin with
\begin{eqnarray}
A=\sum_{j=1}^{2^{m_0-1}} \tilde{A}_j  + \sum_{j=2^{m_0-1}+1}^{2^{m_0}} \tilde{A}_j =: A_0+A_1, \label{Eq:decomp:A}
\end{eqnarray}
where $A_{0}$ and $A_{1}$ are $(g/2)$-extensive.
Due to Theorem~\ref{norm_momentum_generator}, we obtain 
\begin{eqnarray}
M (A,\rho,\tau) &\le  \frac{M (2A_0,\rho,\tau)+M (2A_1,\rho,\tau)}{2}+\frac{3\tau^2}{2} \|[A_0,A_1]\| \nonumber \\
&\le \frac{M (2A_0,\rho,\tau)+M (2A_1,\rho,\tau)}{2}+\frac{3\lambda gN \tau^2}{8}    ,
\end{eqnarray}
where we utilize the inequality $\|[A_0,A_1]\| \le \lambda g N/4$.
For $M (2A_0,\rho,\tau)$ and $M (2A_1,\rho,\tau)$, we apply the same procedure.
We then obtain
\begin{eqnarray}
\fl M (A,\rho,\tau) \le& \frac{M (4 A_{00},\rho,\tau) + M (4 A_{01},\rho,\tau) +M (4A_{10},\rho,\tau) + M (4A_{11},\rho,\tau)}{4}\nonumber \\
\fl  &+ \frac{3\lambda gN \tau^2}{4}  ,
\end{eqnarray}
where we decompose $A_{0}=A_{00}+A_{01}$ and $A_{1}=A_{10}+A_{11}$ in the same way as in Eq.~\eqref{Eq:decomp:A}.
By repeating this process $m_0$ times, we can prove the Corollary~\ref{norm_momentum_generator2}.
$\square$

\subsection{Proof of Theorem~\ref{norm_momentum_generator_localize}}
\label{sec:expE2}

From the assumption of \eqref{a:assump_localize}, we have 
\begin{eqnarray}
\label{rho:ineq:localize}
\rho = \Pi_{|x| \le \sigma}^{(j)}\rho
\end{eqnarray}
for all of $\{A_{j}^\co\}_{j=1}^{\bar{n}}$. 
Now, $\{A_{j}^\co\}_{j=1}^{\bar{n}}$ are given by the summation of non-overlapping local operators.
Hence, on the spectral properties of $\{A_{j}^\co\}_{j=1}^{\bar{n}}$, we obtain the following inequality~\cite{arad2014connecting} for any $q$-local operator $O$:
\begin{eqnarray}
\label{commuting:AKL:bound}
\left \| \Pi_{\ge \epsilon+\Delta  \epsilon}^{(A_{j}^\co)} O \Pi_{\le  \epsilon}^{(A_{j}^\co)} \right\| 
 \cases{
\le \|O\|  \for  \Delta  \epsilon \le  2gq,  \\ 
=0  \for  \Delta  \epsilon > 2gq.}
\end{eqnarray}
where each of $\left\{\Pi^{(A_{j}^\co)}_{\ge  \epsilon}\right\}_{j=1}^{\bar{n}}$ denotes the projection operator onto the eigenspace of $A_{j}^\co$ which is in $[\epsilon,\infty)$.
By using the inequalities~\eqref{rho:ineq:localize} and~\eqref{commuting:AKL:bound}, we have
\begin{eqnarray}
\Pi_{|x| \ge h}^{(A_{j}^\co)} A^{m} \rho =\Pi_{|x| \ge h}^{(A_{j}^\co)} A^{m} \Pi_{|x| \le \sigma}^{(A_{j}^\co)}\rho =
0  \for  h-\sigma > 2gkm,
\end{eqnarray}
where we use the fact that $A^{m}$ is at most $(mk)$-local.
This equality gives 
\begin{eqnarray}
\label{Eq:tr_Am:sequence}
\tr \left ( A^{m} \rho \right) =\tr \left( \mathcal{A}_{m}\mathcal{A}_{m-1}\cdots \mathcal{A}_{2}\mathcal{A}_{1}\rho\right) 
\end{eqnarray}
with
\begin{eqnarray}
\mathcal{A}_{s} = \frac{1}{\bar{n}} \sum_{j=1}^{\bar{n}} \Pi_{|x| \le \sigma + 2gk(s-1)}^{(A_{j}^\co)} A_{j}^\co
\end{eqnarray}
 for $s=1,2,\ldots m$. Note that 
 \begin{eqnarray}
 \label{ineqa:norm:As}
\fl \| \mathcal{A}_{s}\| \le \frac{1}{\bar{n}} \sum_{j=1}^{\bar{n}}\left \| \Pi_{|x| \le \sigma + 2gk(s-1)}^{(A_{j}^\co)} A_{j}^\co \right\| \le \frac{\sigma + 2gk(s-1)}{\bar{n}} \sum_{j=1}^{\bar{n}} = \sigma + 2gk(s-1).
\end{eqnarray}
 
From Eq.~\eqref{Eq:tr_Am:sequence} and the inequality~\eqref{ineqa:norm:As}, we obtain
\begin{eqnarray}
\tr \left ( A^{m} \rho \right)  \le  \lambda^{n } [(m-1) +\sigma/\lambda] [(m-2) +\sigma/\lambda]  \cdots \sigma/\lambda  .
\end{eqnarray}
We thus arrive at
\begin{eqnarray}
\fl  \tr \left ( e^{\tau A} \rho \right)  &\le \sum_{m=0}^{\infty} \frac{( \lambda \tau )^{m} }{m!} [(m-1) +\sigma/\lambda] [(m-2) +\sigma/\lambda]  \cdots \sigma/\lambda    = (1-\lambda \tau)^{-\sigma/\lambda}
\end{eqnarray}
for $0 \le \tau\le \lambda$.  The last equality comes from the taylor expansion of $(1-x)^{-s}= \sum_{n=0}^{\infty}\frac{x^{n}}{n!} s (s+1) \cdots (s+n-1)$.
This completes the proof.
$\square$

\subsection{Proof of Corollary~\ref{norm_momentum_generator_appli}}
We can obtain the inequality~\eqref{general_ver_AKL}  immediately from Lemma~\ref{decomposition} and  Theorem~\ref{norm_momentum_generator_localize}.
From Lemma~\ref{decomposition}, the Hamiltonian can be decomposed into the $k$-local and $(gk)$-extensive operators in the form of Eq.~\eqref{construction_Commuting_Hm}, say $\{H_{j}^{\co}\}_{j=1}^{\bar{n}}$, where $H_{j}^{\co}\ket{\Omega}=0$ for  $j=1,2,3,\ldots,\bar{n}$ because $H$ is frustration free.
From the inequality~\eqref{commuting:AKL:bound} and $q$-locality of the operator $A$, the energy distribution of $A \ket{\Omega}$ with respect to each of the Hamiltonians $\{H_{j}^{\co}\}_{j=1}^{\bar{n}}$ is completely localized at most with the width of $2gkq$, that is, $\sigma=2gkq$.
Then, the inequality~\eqref{norm_momentum_generator2_localize} in Theorem~\ref{norm_momentum_generator_localize} gives
\begin{eqnarray}
\frac{\| \Pi^{H}_{\ge x} A \ket{\Omega}\|^{2} }{\|A \ket{\Omega}\|^{2}}\le   \left(\frac{x}{2gkq} \right)^{q/k} \exp \left( - \frac{x - 2 gkq}{k \lambda} \right) , \for x\ge 2gkq , 
\label{general_ver_AKL_bef}
\end{eqnarray}
which reduces to the inequality~\eqref{general_ver_AKL}. Notice that the Hamiltonian is now $(gk)$-extensive instead of $g$-extensive, and hence the parameter $\lambda$ in the inequality~\eqref{norm_momentum_generator2_localize} should be replaced by $k\lambda$.
$\square$

\subsection{Proof of Lemma~\ref{Chernoff_finite_dim}}
\label{sec:Ht}

The proof is immediately given by applying Corollary~\ref{norm_momentum_generator2} and the inequality~\eqref{Chernoff_ineq_moment}.
First, from Corollary~\ref{norm_momentum_generator2}, we have, 
\begin{eqnarray}
M (A ,\rho_\Prod,\tau) &\le  \frac{1}{2^{m_0}} \sum_{j=1}^{\bar{n}} M \left ( 2^{m_0} A^\co_j/\bar{n} ,\rho_\Prod,\tau \right) + \frac{3\lambda gN \tau^2}{8} \lceil \log_2 \bar{n} \rceil   . 
\label{moment:ineq:finite:dim:proof1}
\end{eqnarray}
By replacing $\tau$ with $ \frac{2^{m_{0}}}{\bar{n}}g \tau $ in the inequality~\eqref{Chernoff_ineq_moment}, we obtain  
\begin{eqnarray}
M \left ( 2^{m_0} A^\co_j/\bar{n} ,\rho_\Prod,\tau \right)  \le  \frac{C}{k} N  \tau^{2} \left( \frac{2^{m_0}g }{\bar{n}} \right)^{2}
\end{eqnarray}
for $j=1,2,\ldots,\bar{n}$.
This reduces the inequality~\eqref{moment:ineq:finite:dim:proof1} to
\begin{eqnarray}
M (A,\rho_\Prod,\tau) &\le \frac{1}{2^{m_0}}  \sum_{j=1}^{\bar{n}} \frac{C}{k}  N \left( \frac{2^{m_0}g \tau}{\bar{n}}\right)^2 +\frac{3\lambda gN \tau^2}{8} \lceil \log_2 \bar{n} \rceil     \nonumber \\
&\le N\tau^2 \left(\frac{2g^{2} C}{k} +\frac{3\lambda g}{8} \lceil \log_2 \bar{n}\rceil \right ),
\end{eqnarray}
where we use  $2^{m_0} \le 2\bar{n}$. We thus prove Lemma~\ref{Chernoff_finite_dim}.
$\square$

\subsection{Proof of Theorem~\ref{Chernoff_infinite_dim}}

For the proof, we calculate the moment $\langle A^m \rangle$ separately. 
The main idea for the calculations is to decompose $\langle A^m \rangle$ as follows. 
We begin with $m=1$ and $m=2$, and consider the decomposition of 
\begin{eqnarray}
&\langle A^1 \rangle= \frac{1}{\bar{n}}\sum_{j=1}^{\bar{n}} \langle A^{\co}_{j} \rangle  ,\nonumber \\
&\langle A^2\rangle= \frac{1}{\bar{n}^{2}} \sum_{i,j=1}^{\bar{n}} \langle A_i^{\co} A_j^{\co} \rangle =   \frac{1}{\bar{n}^{2}} \sum_{i<j} \langle (A^{\co}_{i}+A^{\co}_{j})^{2} \rangle -\frac{\bar{n}-2}{\bar{n}^{2}} \sum_{j=1}^{\bar{n}} 
\langle (A^{\co}_{j})^{2} \rangle  .
\end{eqnarray}
The point of these decomposition is that we only have to calculate a set of the moment functions which are constructed from at most two operators. 
For example,  $A^{\co}_{i}+A^{\co}_{j}$ is corresponding to the case of $\bar{n}=2$, and hence the upper bound of 
$\langle (A^{\co}_{i}+A^{\co}_{j})^{2}\rangle$ can be efficiently estimated from the inequality~\eqref{ineq:distribution_gaussian_chernoff}.

In the same way, we aim to decompose $\langle A^m\rangle$ so that we may have to calculate a set of the moment functions which are constructed from at most $m$ operators. 
For this purpose, we consider the following decomposition:
 \begin{eqnarray}
A^m = \sum_{j=0}^{m-1} \theta_j S_{m-j}.
\label{Decomp_H_m_moment}
\end{eqnarray}
Here, $\{\theta_j\}_{j=0}^{m-1}$ are integers and 
\begin{eqnarray}
S_{m-j} :=\frac{1}{\bar{n}^{m}} \sum_{|\Xi|=m-j} |\Xi|^mA_{\Xi}^{m}, \quad A_{\Xi} :=\frac{1}{|\Xi|} \sum_{i\in \Xi} A_i^{\co} ,
\end{eqnarray}
where $\Xi$ is a positive-integer set of size $|\Xi|$ which are picked up from $\{1,2,\ldots,\bar{n}\}$, e.g. $\Xi=\{i_{1},i_{2},i_{3},\ldots,i_{|\Xi|}\}$. 
In~\ref{appendix_A}, we show that the integers $\{\theta_j\}_{j=0}^{m-1}$ are given by
\begin{eqnarray}
\theta_j= (-1)^j \binom{\bar{n}-m+j-1}{j}   .  \label{equality decomp coefficient moment}
\end{eqnarray}
By using this decomposition, we obtain
 \begin{eqnarray}
 \label{sum:calculation:Am}
|\langle A^m \rangle| &=\left| \sum_{j=0}^{m-1} \frac{\theta_j}{\bar{n}^m}  \sum_{|\Xi|=m-j}|\Xi|^m  \left\langle A_{\Xi}^{m} \right\rangle \right| \le \sum_{j=0}^{m-1} \frac{|\theta_j|}{\bar{n}^m}    \sum_{|\Xi|=m-j} |\Xi|^m \left| \left\langle A_{\Xi}^{m} \right\rangle \right|.
\end{eqnarray}
Here, the operator $A_{\Xi}$ is a summation of $|\Xi|$ operators, and hence the inequality~\eqref{ineq:distribution_gaussian_chernoff} implies that $\tr \left( \Pi^{A_\Xi}_{\ge  x}\rho_{\Prod}   \right) $ decays as $e^{-{\rm const}. \frac{x^2}{N \log |\Xi|}}$, namely the gaussian decay with the variance $\orderof{N\log |\Xi|}$.
We thus obtain
 \begin{eqnarray}
  \label{sum:calculation:Am_decay}
\fl  | \langle A_{\Xi}^{m} \rangle |&\le c_{1} \Gamma \left( \frac{|\Xi|+1}{2}\right) (c_{2} N \log |\Xi|)^{m/2} \le
c_{1} \Gamma \left( \frac{m+1}{2}\right) (c_{2} N \log m)^{m/2}  ,
\end{eqnarray}
where $\Gamma(x)$ is the gamma function, and $c_1$ and $c_2$ are constants of $\orderof{1}$ which depend on $g$, $k$, and $C$.
Note that  $A_{\Xi}$ is now the summation of at most $m$ operators ($|\Xi| \le m$).

The remaining problem is to count the number of summands in Eq.~\eqref{sum:calculation:Am}.
First, the number of the sets $\Xi$ which satisfy $|\Xi|=m-j$ is given by $\binom{\bar{n}}{m-j}$. 
Hence, the total number of summation in Eq.~\eqref{Decomp_H_m_moment} is
\begin{eqnarray}
 \label{sum:calculation:Am_summand}
\sum_{j=0}^{m-1}| \theta_j| \binom{\bar{n}}{m-j} &=\sum_{j=0}^{m-1} \binom{\bar{n}-m+j-1}{j} \binom{\bar{n}}{m-j}   \nonumber \\
&=\sum_{j=0}^{m-1} \frac{(\bar{n}-m+j-1)!}{j! (\bar{n}-m-1)!} \cdot\frac{\bar{n}!}{(m-j)! (\bar{n}-m+j)!}  \nonumber \\
&= \sum_{j=0}^{m-1} \frac{\bar{n}-m}{\bar{n}-m+j}  \frac{1}{j! (\bar{n}-m)!} \cdot\frac{\bar{n}!}{(m-j)!} \nonumber \\
&\le \binom{\bar{n}}{m}  \sum_{j=0}^{m-1} \binom{m}{j} \le 2^m  \binom{\bar{n}}{m} .
\end{eqnarray}

By combining the inequalities~\eqref{sum:calculation:Am}, \eqref{sum:calculation:Am_decay} and \eqref{sum:calculation:Am_summand}, 
we  obtain
\begin{eqnarray}
\label{moment_ineq_upp}
| \langle A^m \rangle | &\le  \frac{2^m}{\bar{n}^{m}} \binom{\bar{n}}{m}  c_{1} \Gamma \left( \frac{m+1}{2}\right) (c_{2} N \log m)^{m/2}    \nonumber \\
&\le c_1 (2N c_2 m \log m)^{m/2},
\end{eqnarray}
where we use the inequalities $\binom{\bar{n}}{m}\le \bar{n}^m$ and $\Gamma \left( \frac{m+1}{2}\right) \le (m/2)^{m/2}$.
To connect the inequality~\eqref{moment_ineq_upp} to the distribution function of $A$, we use the following inequality:
\begin{eqnarray}
| \langle A^m \rangle |&\ge x^{m} \tr (\Pi_{\ge x} \rho_{\Prod})  \for x>0,
\end{eqnarray}
which yields with \eqref{moment_ineq_upp} 
\begin{eqnarray}
\tr (\Pi_{\ge x} \rho_{\Prod})\le c_1  \left( \frac{2N c_2 m \log m}{x^2} \right)^{m/2} .
\end{eqnarray}
By defining $m_0$ as the minimum integer such that 
\begin{eqnarray}
\label{def:m0:general}
 \frac{2N c_2 m \log m}{x^2} \le \frac{1}{e},
\end{eqnarray}
we obtain 
\begin{eqnarray}
\label{def:m0:general_choice}
\tr (\Pi_{\ge x} \rho_{\Prod})\le c_1  e^{-m_0/2} .
\end{eqnarray}
Now, the integer $m_0$ satisfying \eqref{def:m0:general} has the value of $\frac{{\rm const}. x^2}{N \log (x^2/N)}$ and the inequality~\eqref{def:m0:general_choice} reduces to \eqref{ineq:Chernoff_infinite_dim}. 
We thus prove the theorem.
$\square$

\section{Summary and future work}
\label{sec:conc}

In the first half of this paper, we have proved general theorem on the moment generating function, 
which connects a set of moment generating function $\{M(\rho, \tau, A_i)\}_{i=1}^{\bar{n}}$ to $M(\rho, \tau, \sum_{i=1}^{\bar{n}} A_i)$. 
The crucial point of this statement is to restrict ourselves to the few-body operators instead of the arbitrary operators. 
This result is quite general in the sense that we do not need any assumption on the quantum state $\rho$ and the few-body operators are most general observables when we analyze the realistic quantum many-body systems.  
Also, experimentally, this analysis may be helpful to infer probability distributions for other eigenbases which are difficult to experimentally measure; it is often the case where we can prepare projective measurements only onto simple eigenbases.
As one of the applications, we utilize this upper bound to generalize the Chernoff inequality so that it may be applied to more general observables beyond summation of single-site operators as in Fig.~\ref{fig:Chernoff_ineq}. 

On the other hand, we have left several open problems. 
First, in Corollary~\ref{norm_momentum_generator2}, the error term in \eqref{norm_momentum_generator_thm2} is proportional to $\log \bar{n}$. 
This term obstacles us in Theorem~\ref{Chernoff_infinite_dim} to generalize the Chernoff inequality for generic few-body operators in the complete way, i.e., in the strict Gaussian form instead of the quasi-Gaussian form.
So far, we have not clarified whether this estimation is qualitatively optimal or not. 
The difficulty comes from that the scaling of $\log \bar{n}$ is too subtle to observe in numerical ways.

Second, as an important future direction, can we apply the present techniques to analyze more general quantum states?
Our approach gives strong statements on the Chernoff inequality, whereas the range of application is now restricted to the cases where the quantum state $\rho$ is given by the product states or the short-range entangled states.
One of the most prominent classes are the gapped ground states, or equivalently ground states in non-critical phases. 
In such systems, the Chernoff-type inequality has been also proved~\cite{Ref:LR} but in a weaker way as $e^{-|x|/\orderof{\sqrt{N}}}$ instead of the Gaussian decay as $e^{-x^2/\orderof{N}}$. 
This kind of the probability-distribution analysis provides us useful information in constructing the approximate ground states projection, which has been 
a backbone in recent ground states' analyses~\cite{Ref:LR,ref:Arad-AL13,arad2016rigorous}. 
It is a quite intriguing problem weather we can refine the weak Chernoff inequality for the gapped ground states to the Gaussian form.

\section*{Acknowledgments}

This work was supported by World Premier International Research Center Initiative (WPI), Mext, Japan.   
TK also acknowledges the support from JSPS grant no. 2611111.

\bibliographystyle{iopart-num}

{~}

\bibliography{edist2}

\appendix

\section{Derivation of the equality~\eqref{equality decomp coefficient moment}}  \label{appendix_A}

For the derivation, we need to subtract the overcounting terms. 
To make the point clear, we begin with the case of $m=2$:
\begin{eqnarray}
A^{2}= \left(\sum_{j=1}^{\bar{n}} A^{\co}_{j}\right)^{2} .
\end{eqnarray}
In this case, in Eq.~\eqref{Decomp_H_m_moment}, we have
\begin{eqnarray}
S_0= \sum_{i<j} (A^{\co}_{i}+A^{\co}_{j})^{2} ,\quad S_1=\sum_{j=1}^{\bar{n}}(A^{\co}_{j})^{2} .
\end{eqnarray}
Then, in $S_{0}$, the terms of $A^{\co}_{i}A^{\co}_{j}$ ($i\neq j$) are not overcounting, but 
the terms of $A^{\co}_{i}A^{\co}_{i}$ are counted $(\bar{n}-1)$ times; that is, we overcount the terms $(n-2)$ times and have to subtract $(\bar{n}-2) S_1$.
We thus obtain $A^2= S_0 - (\bar{n}-2) S_1$, namely $\theta_0=1$ and $\theta_1=-(\bar{n}-2)$.

In the same way, by subtracting the overcounting terms, we reach the following recurrence equation for general $m$:
\begin{eqnarray}
\theta_{j_0}=- \sum_{j=0}^{j_0-1} \theta_j \binom{\bar{n}-m+j_0}{j_0-j}  +1 \for  0 \le j_0 \le m-1
\end{eqnarray}
with $\theta_0=1$. 
We obtain Eq.~\eqref{equality decomp coefficient moment} by solving this equation.
We prove it by the inductive method.
For $j=0$, it is clear that $\theta_{0}=1$, and we assume that the equality is true for $j \le j_{0}-1$.
Then, we obtain 
\begin{eqnarray}
\theta_{j_0}=- \sum_{j=0}^{j_0-1}(-1)^{j} \binom{\bar{n}-m+j-1}{j} \binom{\bar{n}-m+j_0}{j_0-j} +1.
\end{eqnarray}
Our task is to show the equality
\begin{eqnarray}
\fl  -\sum_{j=0}^{j_0-1}(-1)^{j} \binom{\bar{n}-m+j-1}{j} \binom{\bar{n}-m+j_0}{j_0-j}  +1= (-1)^{j_0} \binom{\bar{n}-m+j_0-1}{j_0} ,
\label{induction proof binom eq}
\end{eqnarray}
which gives us the equality~\eqref{equality decomp coefficient moment}. 

For the derivation of Eq.~\eqref{induction proof binom eq}, we use the inductive method again.
First, by denoting $M=\bar{n}-m$, the inequality reduces to
\begin{eqnarray}
-\sum_{j=0}^{j_0-1}(-1)^{j} \binom{M+j-1}{j} \binom{M+j_0}{j_0-j}  +1= (-1)^{j_0} \binom{M+j_0-1}{j_0}    .
\end{eqnarray}
For $j_{0}=1$, we have 
\begin{eqnarray}
- \binom{M-1}{0} \binom{M+1}{1}  +1=  -M = (-1)^{1} \binom{M+1-1}{1}  ,
\end{eqnarray}
and hence the equality is true.
We then assume that the equality is true for the case of $j_{0}$ and prove the case of $j_{0}+1$.
We begin with the decomposition of
\begin{eqnarray}
\label{appendix:eq:summation}
&-\sum_{j=0}^{j_0}(-1)^{j} \binom{M+j-1}{j} \binom{M+j_0+1}{j_0+1-j}  +1\nonumber \\
=& -\sum_{j=0}^{j_0-1} (-1)^{j} \binom{M+j-1}{m} \binom{M+1+j_0}{j_0-j+1}  +1 \nonumber \\
&  -(-1)^{j_0} \binom{M+j_0-1}{j_0} \binom{M+j_0+1}{j_0+1-j_0} .
\end{eqnarray}

For the calculation of the first term in Eq.~\eqref{appendix:eq:summation}, we utilize the Pascal's rule as
\begin{eqnarray}
\fl \binom{M+j_0+1}{j_0-j+1}  &= \binom{M+j +(j_0-j+1)}{j_0-j+1} = \binom{M+j_0}{j_0-j}  + \binom{M+j_0}{j_0-j+1} ,
\end{eqnarray}
which yields
\begin{eqnarray}
\label{appendix:eq:summation1}
\fl&-\sum_{j=0}^{j_0-1} (-1)^{j} \binom{M+j-1}{j} \binom{M+1+j_0}{j_0-j+1}  +1\nonumber \\
\fl=&-\sum_{j=0}^{j_0-1} (-1)^{j} \binom{M+j-1}{j} \left[ \binom{M+j_0}{j_0-j} + \binom{M+j_0}{j_0-j+1} \right]  +1 \nonumber \\
\fl=&(-1)^{j_0}\binom{M+j_0-1}{j_0} -\sum_{j=0}^{j_0-1} (-1)^{j} \frac{(M+j-1)!}{j!(M-1)!} \cdot \frac{(M+j_0)!}{(j_0-j+1)!(M+j-1)!}  \nonumber\\
\fl=& (-1)^{j_0}\binom{M+j_0-1}{j_0} -\sum_{j=0}^{j_0-1} (-1)^{j} \binom{j_0+1}{j}  \binom{M+j_0}{j_0+1} \nonumber \\
\fl=& (-1)^{j_0}\binom{M+j_0-1}{j_0} + (-1)^{j_0} \binom{j_0+1}{j_0}  \binom{M+j_0}{j_0+1}+ (-1)^{j_0+1}  \binom{M+j_0}{j_0+1}  \nonumber \\
\fl=&(-1)^{j_0}\binom{M+j_0-1}{j_0} + (-1)^{j_0} j_0  \binom{M+j_0}{j_0+1} .
\end{eqnarray}
The second term in Eq.~\eqref{appendix:eq:summation} is also given by
\begin{eqnarray}
\label{appendix:eq:summation2}
 \fl -(-1)^{j_0} \binom{M+j_0-1}{j_0} \binom{M+j_0+1}{j_0+1-j_0}=-(-1)^{j_0} \binom{M+j_0-1}{j_0} (M+j_0+1).
\end{eqnarray}
By combining the above equalities~\eqref{appendix:eq:summation1} and \eqref{appendix:eq:summation2} in Eq.~\eqref{appendix:eq:summation}, we arrive at
\begin{eqnarray}
\fl &-\sum_{j=0}^{j_0}(-1)^{m} \binom{M+j-1}{j}\binom{M+j_0+1}{j_0+1-j}+1\nonumber \\
\fl =&(-1)^{j_0}\binom{M+j_0-1}{j_0} + (-1)^{j_0}j_0\binom{M+j_0}{j_0+1}-(-1)^{j_0}\binom{M+j_0-1}{j_0}(M+j_0+1)\nonumber \\
\fl =& (-1)^{j_0}j_0\binom{M+j_0}{j_0+1} -(-1)^{j_0}\binom{M+j_0-1}{j_0}(M+j_0)=(-1)^{j_0+1}\binom{M+j_0}{j_0+1} ,
\end{eqnarray}
where we use the equality
\begin{eqnarray}
\fl (-1)^{j_0}\binom{M+j_0-1}{j_0}(M+j_0)&=(-1)^{j_0} \frac{(M+j_0-1)!}{j_0!(M-1)!} (M+j_0)  \nonumber \\
\fl  &=(-1)^{j_0} (j_0+1) \frac{(M+j_0)!}{(j_0+1)!(M-1)!}.
\end{eqnarray}
We thus prove the equality~\eqref{induction proof binom eq} for the case of $j_0+1$. 

This completes the proof. $\square$


\end{document}